\documentclass[twocolumn]{aastex701} 
\usepackage{soul}
 \usepackage{pifont}

\newcommand{\nfwhmlo}{$130$ km s$^{-1}$}
\newcommand{\nfwhmme}{$185$ km s$^{-1}$}
\newcommand{\nfwhmhi}{$250$ km s$^{-1}$}

\begin{document}

\title{Constraints on the Gas Geometry Surrounding Little Red Dots through Narrow-Line Diagnostics}

\author[orcid=0000-0003-0780-9526]{Visal Sok}
\affiliation{Department of Astrophysical \& Planetary Sciences, University of Colorado, 2000 Colorado Ave, Boulder, CO 80309, USA}
\email[show]{visal.sok@colorado.edu}  
\author[]{Erica J. Nelson}
\affiliation{Department of Astrophysical \& Planetary Sciences, University of Colorado, 2000 Colorado Ave, Boulder, CO 80309, USA}
\email[no]{}    
\author[orcid=0000-0003-0936-8488]{Mitchell C. Begelman}
\affiliation{JILA, University of Colorado and National Institute of Standards and Technology, 440 UCB, Boulder, CO 80309, USA}
\email[no]{}  
\author[orcid=0000-0003-3903-0373]{Jason Dexter}
\affiliation{JILA, University of Colorado and National Institute of Standards and Technology, 440 UCB, Boulder, CO 80309, USA}
\affiliation{Department of Astrophysical \& Planetary Sciences, University of Colorado, 2000 Colorado Ave, Boulder, CO 80309, USA}
\email[no]{}  
\author[orcid=0000-0003-2388-8172]{Francesco D'Eugenio}
\affiliation{Kavli Institute for Cosmology, University of Cambridge, Madingley Road,
Cambridge CB3 0HA, UK}
\affiliation{Cavendish Laboratory -- Astrophysics Group, University of Cambridge, 19
JJ Thomson Avenue, Cambridge CB3 0HE, UK}
\email[no]{}  
\author[orcid=0000-0002-5612-3427]{Jenny E. Greene}
\affiliation{Department of Astrophysical Sciences, Princeton University, Princeton, NJ 08544, USA}
\email[no]{}    
\author[orcid=0000-0001-6755-1315]{Joel Leja}
\affiliation{Department of Astronomy \& Astrophysics, The Pennsylvania State University, University Park, PA 16802, USA}
\affiliation{Institute for Computational \& Data Sciences, The Pennsylvania State University, University Park, PA 16802, USA}
\affiliation{Institute for Gravitation and the Cosmos, The Pennsylvania State University, University Park, PA 16802, USA}
\email[no]{}  
\author[0000-0001-7160-3632]{Katherine E. Whitaker}
\affiliation{Department of Astronomy, University of Massachusetts, Amherst, MA 01003, USA}
\affiliation{Cosmic Dawn Center (DAWN), Copenhagen, Denmark}
\email[no]{}    
\author[orcid=0000-0002-8651-9879]{Andrew J. Bunker}
\affiliation{Department of Physics, University of Oxford, Denys Wilkinson Building, Keble Road, Oxford OX1 3RH, U.K.}
\email[no]{}  
\author[0000-0003-4528-5639]{Pablo G. P\'erez-Gonz\'alez}
\affiliation{Centro de Astrobiolog\'ia (CAB), CSIC-INTA, Ctra. de Ajalvir km 4, Torrej\'on de Ardoz, E-28850, Madrid, Spain}
\email[no]{}  
\author[orcid=0000-0002-5104-8245]{Pierluigi Rinaldi}
\affiliation{Space Telescope Science Institute, 3700 San Martin Drive, Baltimore, Maryland 21218, USA}
\email[no]{}    
\author[orcid=0000-0001-5586-6950]{Alberto Torralba}
\affiliation{Institute of Science and Technology Austria (ISTA), Am Campus 1, 3400 Klosterneuburg, Austria}
\email[no]{}    
\author[0000-0003-4891-0794]{Hannah \"Ubler}
\affiliation{Max-Planck-Institut f\"ur extraterrestrische Physik (MPE), Gie\ss enbachstra\ss e 1, 85748 Garching, Germany}
\email[no]{}   

\begin{abstract}

Little Red Dots (LRDs) are a recently identified population of high-redshift sources, with a common interpretation being accreting black holes embedded within a spherical, optically thick gas envelope. Within this framework, 
some models propose that the continuum arises from the dense-gas envelope, where hard ionizing radiation from the central engine is reprocessed into a stellar-like photosphere with an effective temperature of $\sim$5000 K.
This implies that both the UV continuum and narrow-line emission are then powered by the host galaxy rather than an exposed central engine. 
To test whether this is consistent with the observed narrow-line ratios, we analyze multiple line diagnostics for a sample of $\sim$20 LRDs with high signal-to-noise NIRSpec grating spectra. We find that at least 40\% of the LRDs have line ratios pointing toward high ionization parameter and electron temperature, with a further 15\% also falling in the AGN regime for the O\textsc{i}/H$\alpha$ diagnostic, indicative of harder ionizing radiation. These line ratios are incompatible with stellar photoionization from a star-forming host alone. This suggests lower density channels within the gas envelope through which high energy photons can escape and excite the surrounding narrow-line emitting gas. 
At the same time, most LRDs lack strong high-ionization line emission, with He\,\textsc{ii}/H$\beta$ $\lesssim0.1$, consistent with an ionizing spectrum softer than that of a standard AGN. Together, these results disfavour a uniform gas envelope with a covering fraction of unity, and instead point to a more complex geometry that gives rise to anisotropic ionizing radiation. 
\end{abstract}



\section{Introduction} 

One of the most intriguing discoveries from the James Webb Space Telescope (JWST) is a new population of high-redshift objects known as Little Red Dots (LRDs; e.g., \citealt{Labbe2023, Kocevski2023, Greene2024, Matthee2024}), traditionally selected based on their red colours and point-like morphologies in NIRCam long-wavelength imaging. Since their discovery, the physical nature of LRDs remains an open question, and has prompted extensive analysis of whether their emission lines and v-shaped SEDs are driven by dust-reddened AGN, gas-enshrouded supermassive black holes, massive galaxies with old stellar populations, or some combination of these. 

The motivation for LRDs to be dust-reddened active galactic nuclei (AGN) includes distinctive spectral features such as broad emission lines, red optical continua, and strong Balmer breaks. However, LRDs show weak X-ray emission (e.g., \citealt{Ananna2024, Yue2024, Kokubo2025, Sacchi2025}), and lack the mid-infrared hot dust emission associated with a dusty torus (e.g., \citealt{Perez2024, Williams2024, Akins2025, Wang2025_dust, Setton2025_dust}). The red optical continua of LRDs also cannot be explained by attenuation from ISM dust, given the lack of far-infrared emission (e.g., \citealt{Casey2024, Akins2025, Chen2025, Setton2025_dust, Xiao2025}). However, the presence of dust may vary across LRD populations, where the inferred slope between 1-3 microns from stacked MIRI photometry is found to become progressively redder toward LRDs with lower optical-to-UV luminosity ratios \citep{Perez2026}. 
The Balmer break observed in LRDs was also initially attributed to evolved stellar populations, which would imply the early formation of extremely massive galaxies at high-$z$ \citep{Labbe2023}. 
While the rest-UV emission of LRDs is often spatially extended, their rest-optical continuum is extremely compact (e.g., \citealt{Jones2025, Cloonan2026}). If the compact red emission were stellar in origin, the implied mass densities would exceed the theoretical limits for star clusters (e.g., \citealt{Baggen2024, Hopkins2010}). The strength of some observed Balmer breaks is also too extreme to be reproduced by evolved stellar populations alone (e.g., \citealt{Naidu2025, Graaff2025_bh, Deugenio2025}).

A compelling framework to unify these observations describes LRDs as having a significant amount of dense, neutral hydrogen gas surrounding a central engine, where the gas is collisionally excited into the $n=2$ states \citep{Juodzbalis2024, Inayoshi2025}. Such a configuration is predicted in objects such as late-stage quasi-stars (e.g., \citealt{Begelman2006, Begelman2026}) or ``Black Hole Stars” (e.g., \citealt{Naidu2025, Graaff2025_bh}).
The dense gas envelope, with a covering fraction near unity, can successfully reproduce several key spectral features of LRDs. Specifically, a Balmer break is expected due to the absorption of photons blueward of the Balmer limit by hydrogen atoms in the $n=2$ state, while the broad line profiles may result from electron scattering as radiation is reprocessed \citep{Rusakov2026}, or through other broadening mechanisms \citep{Madau2026, Scholtz2026}. The limited detection of X-ray emission and high-ionization lines can also be explained within this model as absorption by the Compton-thick gas (e.g., \citealt{Juodzbalis2024}). However, these may not be universal as some X-ray luminous AGN may also be LRDs \citep{Hviding2026}. 
Other interpretations of LRDs include direct collapse black holes (e.g., \citealt{Pacucci2026, Baggen2026}), supermassive stars (e.g., \citealt{Zwick2025, Nandal2026}), and globular cluster formation \citep{Chisholm2026}.  

While a uniform gas envelope provides a coherent physical framework, the recent detections of Ly$\alpha$ emission in several LRDs \citep{Tang2026, Ji2026} suggest that the gas envelope is clumpy, with low-density channels for photons to escape. Variability in the broad and narrow line fluxes of an LRD between epochs may further point toward an anisotropic gas geometry with covering fraction less than unity \citep{Lambrides2026}. However, such variability may not be universal as no significant variability was detected in the continuum or line fluxes for the TWINKLE sample \citep{Liu2026}. Additional constraints on the gas geometry can also come from the narrow line emissions, where a near-unity covering fraction would imply reprocessed radiation and narrow line ratios consistent with stellar photoionization, and AGN-like ratios would indicate that hard ionizing radiation can escape. This motivates a more systematic investigation of the narrow line properties across the LRD population.

In this paper, we combine new NIRSpec observations of several LRDs with publicly available NIRSpec spectra from the Dawn JWST Archive (DJA) to investigate the origins of their narrow lines. 
We test whether the narrow lines are stellar or AGN in origin by using several independent line diagnostics, and test whether they consistently point to the same conclusion.
These include the new AGN line diagnostics from \cite{Mazzolari2024}, an alternative to the BPT diagram \citep{Baldwin1981} at high redshifts, as well as the OHNO \citep{Backhaus2022} and O1-V087 \citep{Veilleux1987} diagnostics
The paper is structured as follows: \S \ref{sec:data} describes the data and LRD samples in this work. We describe our fitting procedures in \S \ref{sec:methods}. Section \ref{sec:line_ratio} presents the analyses of the narrow lines observed in LRDs. We discuss the implications of our analyses and present our conclusions in \S \ref{sec:discussion}.



\section{Data and LRD Samples} \label{sec:data}

We compile a sample of LRDs from different JWST observations. This includes the final observations from the Cycle-2 JWST program ID 4106 (PIs E. J. Nelson and I. Labb\'e), which were observed in Cycle-3 as compensation for unusable grating exposures due to a `mirror-tilt' event in the original observations. In total, this program includes six LRDs.
We also make use of the observations from the JADES Dark Horse survey \citep{Deugenio2025_DH} from the JWST PID 3215 (PIs D. Eisenstein \& R. Maiolino), which introduced a novel observing strategy using JWST/NIRSpec dense-shutter spectroscopy. These spectroscopic observations originated from failed MSA shorts in Cycle 2 of PID 3125, which were reallocated to Cycle 3, but faced additional delays. The program was subsequently redesigned as a pilot study to showcase a dense-shutter spectroscopy strategy that maximizes the number of observable spectra. Further details of the observations and data reduction are presented in \cite{Deugenio2025_DH}. The Dark Horse sample includes a total of seven LRDs.  

In addition, we utilize all the publicly available NIRSpec spectra on the DAWN JWST Archive (DJA, \citealt{Brammer2025}). We use the catalogue provided by \cite{Graaff2025_unified_family}, consisting of 116 unique LRDs drawn from different surveys, including CANUCS \citep{Sarrouh2026}, CAPERS (PID 6368; PI: Dickinson), CEERS \citep{Finkelstein2025}, JADES \citep{Eisenstein2026, Bunker2024, DEugenio2025_jades, CurtisLake2026}, NEXUS \citep{Shen2024}, NIRSpec GTO-Wide \citep{Maseda2024}, RUBIES \citep{Graaff2025}, and UNCOVER \citep{Bezanson2024, Price2025}. While 106 of the 116 have PRISM spectra available on the DJA, only 59 have NIRSpec grating observations in addition to the PRISM. The data reduction for the spectra on the DJA is described in \cite{Graaff2024} and \cite{Heintz2024}, with one-dimensional spectra from DJA version 4.4 reduced using \texttt{msaexp} \citep{Brammer2023} and optimally extracted \citep{Horne1986}. 

\subsection{LRD Selection}
Since our sample combines LRDs drawn from different surveys, we refit the UV-optical slope following the strategy of \cite{Hviding2025} to ensure consistent selection criteria across all sources. In particular, the LRDs in these analyses are selected based on the v-shaped UV-optical continuum and their point-source morphology in the rest-frame optical.

The UV slope $\beta_\mathrm{UV}$ and optical slope $\beta_\mathrm{opt}$ are determined by independently fitting a power law of the form $f_{\lambda}\propto \lambda^{\beta}$ to two separate wavelength regions on either side of the Balmer limit using chi-squared minimization, where the Balmer limit serves as the pivot wavelength \citep{Setton2025}.
Specifically, we fit $\beta_\mathrm{UV}$ over the rest-frame range 1200\,\AA\ to 3600\,\AA, and $\beta_\mathrm{opt}$ over 4000\,\AA\ to 7000\,\AA.
Where available, we perform this fit using the PRISM spectra, excluding spectral regions contaminated by emission lines. For sources lacking PRISM spectroscopy (i.e., the DarkHorse sample), $\beta_\mathrm{UV}$ and $\beta_\mathrm{opt}$ are instead derived from photometry \citep{Kocevski2025}. In particular, we use available NIRCam broadband filters whose central wavelengths fall within the respective rest-frame UV and optical ranges defined above, requiring at least two filters per region. Following \cite{Kocevski2025}, we do not correct for emission line contamination, but we acknowledge that this may have an effect on the selection \citep{Rinaldi2026}.

Sources with v-shaped continua are those that satisfy the following criteria:

\begin{itemize}
    \item  $\beta_\mathrm{opt} > 0$
    \item  $\beta_\mathrm{UV} < -0.2$
    \item  $\beta_\mathrm{UV} - \beta_\mathrm{opt} > 0.5$
\end{itemize}
Finally, these sources satisfy the compactness criterion, with $f_\mathrm{F444W(0.^{\prime\prime}2)}/f_\mathrm{F444W(0.^{\prime\prime}1)}<1.7$ ( \citealt{Graaff2025_bh}).
While the parent sample consists of 120 LRDs, the analyses presented in \S\ref{sec:line_ratio} are limited to LRDs with available observations in either NIRSpec medium (R$\sim$1000) or high-resolution (R$\sim$2700) gratings, with spectral coverage spanning approximately 3500\,\AA\ to 7000\,\AA, and with robustly detected emission lines (signal-to-noise ratio, S/N, greater than 3). For the latter criterion, the median S/N after the S/N cut is approximately 7 for O\textsc{iii}$\lambda$4363 and 6 for H$\gamma$.The first and second criteria exclude approximately half the parent sample, and a quarter of the parent sample, respectively. The final sample size presented later in the analyses consist of 19 LRDs.


\section{Methodology} \label{sec:methods}

We develop a custom python-based spectral fitting code to robustly constrain emission line properties from the available NIRSpec spectroscopy. The code simultaneously fits a user-specified set of emission lines across multiple NIRSpec dispersers where available, while accounting for instrument and observational effects. 

We model each emission line as a sum of up to four components, consisting of a narrow, broad (double Gaussian), and where applicable, absorption component. In the current implementation, all the narrow lines share a common redshift, and are modelled as a Gaussian. We further model the broad component for the Balmer lines as a double Gaussian, each with a velocity offset. We opt for a double Gaussian as most of the spectra in our sample have insufficient S/N to distinguish between different profile shapes. We note that an exponential profile is proposed as a feature of LRD broad lines \citep{Rusakov2026}, but this profile is not generally found across the population \citep{Scholtz2026}. For LRDs with visually identified Balmer absorption features, we model these features using a standard attenuation model following \cite{Juodzbalis2024}. The transmitted fraction is given by:
\begin{equation}
    f_\lambda = 1 - C_f + C_f e^{-\tau_\lambda}, 
\end{equation}
where $C_f$ is the covering fraction, assumed to be unity, and $\tau_\lambda$ is the optical depth profile. The optical depth profile at each Balmer line is assumed to be Gaussian, with the three free parameters being the optical depth at line centre $\tau_o$, and the velocity shift and dispersion. All Balmer lines share the same velocity shift and dispersion, but each line has an independent $\tau_o$.

The continuum is modelled locally around each emission line by fitting a third-order polynomial to line-free spectral regions, and subtracting it prior to fitting. The emission model is then broadened by the respective NIRSpec line spread function (LSF). 
To properly handle the pixel sampling of the NIRSpec detectors, each model is constructed at a higher spectral resolution before being resampled to match the native resolution of the observed spectrum.

\begin{figure*}[!t]
    \centering
    \includegraphics{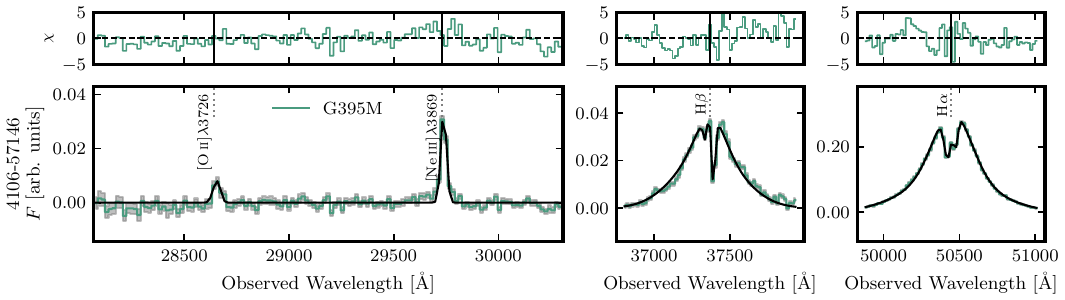}
    \includegraphics{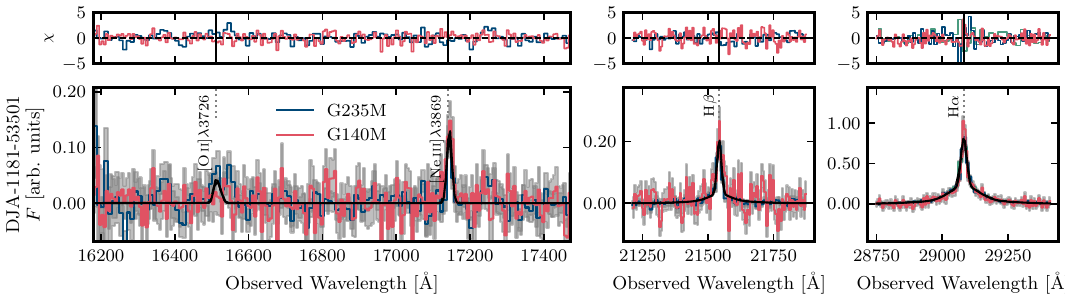}
    \includegraphics{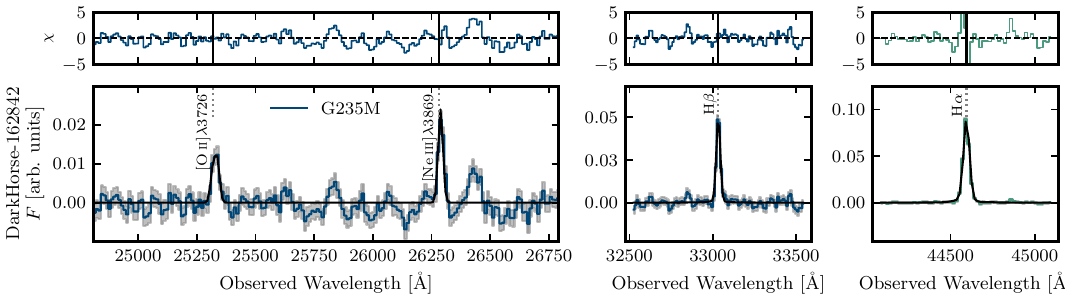}
    \caption{A few examples of the best-fit spectral model ordered by the signal-to-noise ratio of H$\gamma$, showing the regions around key emission lines used in our AGN diagnostic analysis (see \S \ref{sec:line_ratio}). The black line shows the best-fit model, with different colours denoting observed spectra with different dispersers.}
    \label{fig:fit_example}
\end{figure*}

To this end, we only fit the emission lines for LRDs with available medium- and high-resolution grating, and omit those with only PRISM data. We assume a Bayesian framework with flat priors. We sample the posteriors distribution using a Markov-Chain Monte Carlo sampler (\texttt{emcee} \citealt{Mackey2013}), initializing $40\times N_\mathrm{param}$ walkers with 10000 steps. For each free parameter, we consider its 1D marginalized posterior, taking the 50th percentile as the best estimate, and the 16th- to 84th-percentile range as the uncertainties. 
Figure \ref{fig:fit_example} shows a few examples of our simultaneous broad and narrow line fitting. For the following analyses, we only consider robustly detected lines, defined to have a signal-to-noise ratio (S/N) greater than 3. 

\section{Properties of LRDs}

\subsection{Narrow-Line Diagnostics} \label{sec:line_ratio}

A dense gas envelope with a high covering fraction implies the envelope is optically thick to ionizing radiation emitted from the central black hole (e.g., \citealt{Juodzbalis2024}). In this scenario, the narrow lines are suggested to be the result of photoionization from nearby young massive stars. Narrow-line diagnostics therefore offer a window into the ionization conditions of the surrounding gas, and in particular whether it is photoionized by stars or an AGN. 

In this section, we make use of the $[\text{O\,\textsc{iii}}]\lambda4363$ auroral line, OHNO, and O1-VO87 diagnostics to provide insights into the ionizing source. While we adopt demarcation lines from the literature to separate AGN from star-forming regions, we also show photoionization model grids for star formation using \textsc{cloudy} v25 \citep{Gunasekera2025} for reference. 
The incident ionizing SEDs are from BPASS \citep{Eldridge2017PASA...34...58E}, assuming a Chabrier IMF with an upper mass limit of 300 $M_\odot$. We assume a constant star formation history over 10 Myr and match the stellar metallicity to the gas-phase metallicity. We compute grids spanning ionization parameters over $\log U \in [-3, -1]$ and the gas-phase metallicities over $Z \in [0.001, 0.02]$, with the hydrogen density fixed at $10^3~\mathrm{cm}^{-3}$. 

\subsubsection{Auroral Line Diagnostics} \label{sec:auroral_line}

We use the line diagnostics presented by \cite{Mazzolari2024}, which were proposed as an alternative to the Baldwin, Phillips and Terlevich diagram \citep{Baldwin1981}.
At higher redshifts, the BPT diagram is no longer effective (e.g., \citealt{Shapley2004, Kewley2013, Ubler2023, Maiolino2024}), as lower stellar metallicity pushes star-forming galaxies towards higher ionization in [O\textsc{iii}]/H$\beta$, while several effects (lower N/O, high ionization, lack of X-rays) may be pushing AGN towards low [N\textsc{ii}]/H$\alpha$. The result is that the two populations end up occupying the same region of the BPT diagram, becoming indistinguishable. This degeneracy may be mitigated with the inclusion of velocity dispersion \citep{Zhu2025}. 

Regardless, the \citet{Mazzolari2024} diagnostics are instead built around the  
$[\text{O\,\textsc{iii}}]\lambda4363$ auroral line, which is sensitive to electron temperature, motivated by earlier works that attributed elevated values of $\text{O3H}\gamma \equiv [\text{O\,\textsc{iii}}]\lambda4363/\text{H}\gamma$ to AGN activities (e.g., \citealt{Ubler2024}). Combined with ratios such as
$\text{O33} \equiv [\text{O\,\textsc{iii}}]\lambda5007/[\text{O\,\textsc{iii}}]\lambda4363$, which traces the electron temperature, and $\text{Ne3O2}$, which traces the ionization parameter, these diagnostics compare the heating of the gas relative to the ionization parameter, where AGN radiation is suggested to be more effective at heating the gas than star formation at a given ionization parameter.

\begin{figure*}
    \centering
    \includegraphics{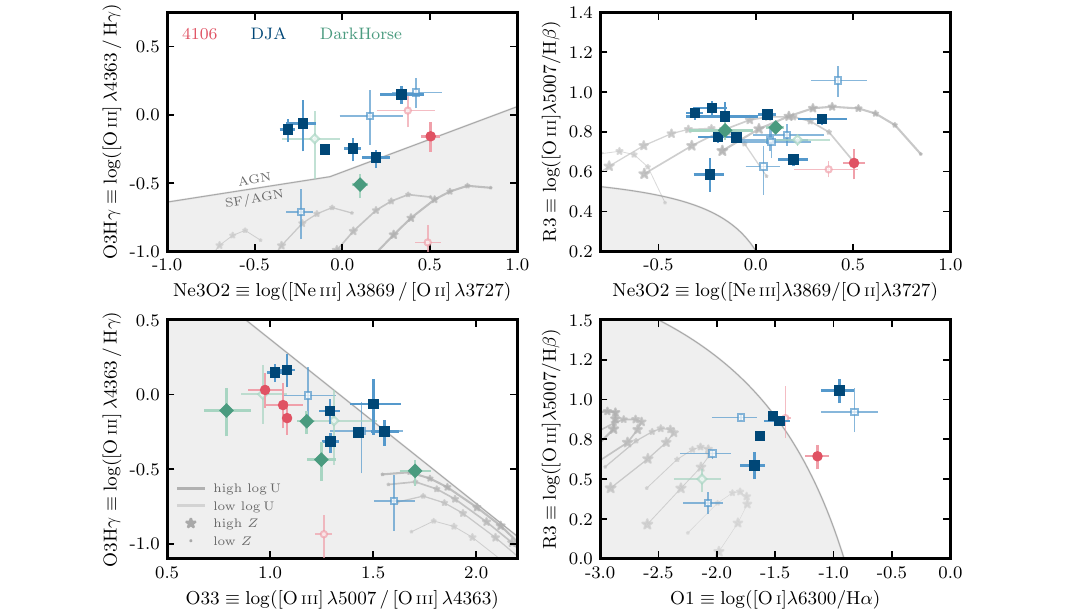}
    \caption{We show the location of LRDs in different narrow-line diagnostic diagrams. The solid markers show the line ratios with S/N greater than 3, while the open markers show line ratios with S/N greater than 2 but less than 3. The shaded regions show the star-forming locus as defined in the literature. We also show stellar photoionization grids based on \textsc{cloudy} modelling for different ionizing parameter ($\log U \in [-3, -1]$) and gas-phase metallicity ($Z \in [0.001, 0.02]$). 
    }
    \label{fig:line_diagnostic}
\end{figure*}

Figure~\ref{fig:line_diagnostic} shows the O3H$\gamma$--Ne3O2 and O3H$\gamma$--O33 
diagrams, with demarcation lines derived from empirical observations and 
\textsc{cloudy} photoionization modelling \citep{Mazzolari2024}. The combined O3H$\gamma$ and O33 ratio is insensitive to dust reddening effects, making it a robust tracer of the electron temperature \citep{Brinchmann2023}. For most plausible densities ($n_e \lesssim 10^5 ~cm^{-3}$), the O33 ratio probes $T_e(\mathrm{O^{++}})$, the temperature of the [O\textsc{iii}] emitting gas, with lower ratios indicating higher temperatures.
 Within this diagram, we find that most LRDs lie within the SF--AGN composite region. Assuming an electron density of $n_e = 10^3~\mathrm{cm}^{-3}$ (typical for star-forming galaxies at high redshift; \citealt{Isobe2023}), we infer $T_e(\mathrm{O^{++}})$ using \textsc{pyneb} \citep{Luridiana2015}. We find that the majority of our sample has electron temperatures of around $T_e\sim20000~\text{K}$, though a few LRDs exceed $T_e>30000~\text{K}$. High electron temperatures ($T_e > 30000~\mathrm{K}$) have been suggested in other LRDs such as CANUCS-LRD-$z$8.6 \citep{Tripodi2025}, and attributed to AGN heating. However, we caution this interpretation may be complicated by the critical density of the [O\,\textsc{iii}] nebular line, with $n_\mathrm{crit} \sim 10^6~\mathrm{cm}^{-3}$, above which O33 becomes less sensitive to temperature and increasingly sensitive to density. Such high densities have been inferred in the LRD candidate ``Irony" by \citet{Deugenio2025}. It should also be noted that the parameter spaces occupied by LRDs are not covered by current photoionization models of AGN and star-formation \citep{Mazzolari2024}, which may in part reflect this density discrepancy.  


Within the O3H$\gamma$--Ne3O2 diagram, we find that the majority of LRDs with robust detections in all relevant lines lie above the demarcation line. The inferred higher electron temperatures cannot be reproduced by stellar photoionization models across a range of ionization parameter and metallicity. However, at a given ionization parameter, a harder intrinsic spectrum heats the gas to a higher temperature than stellar photoionization, suggesting that the gas here is subjected to AGN photoionization. 
We also note that most of these LRDs also fall within the AGN locus based on the demarcation line proposed in \cite{Backhaus2025}, who explored the empirical redshift evolution of this diagnostic. 
Taken together, the extreme electron temperatures inferred from O33 at the given ionization conditions traced by Ne3O2 indicate that LRDs occupy a regime that is difficult to reproduce with standard stellar photoionization alone -- even after allowing for both binaries and very massive stars (100-300 $M_\odot$).


\subsubsection{OHNO Diagnostic}
The OHNO diagnostic is a widely used diagnostic for high-redshift galaxies (e.g., \citealt{Backhaus2022, Backhaus2023}). Similar to the the O3H$\gamma$--Ne3O2 diagram, the OHNO diagnostic also relies on Ne3O2, but incorporates $\mathrm{R3} \equiv [\text{O\,\textsc{iii}}]\,\lambda5007 / \mathrm{H}\beta$.

All of the LRDs in our sample lie above the demarcation line proposed in \cite{Backhaus2022} for galaxies $z\sim1.5$. 
To account for the empirical redshift evolution of both R3 and Ne3O2 \citep{Backhaus2024}, we also explore the demarcation line extrapolated to $z\sim5$, corresponding to the median redshift of our LRD sample. Even with this correction, most LRDs remain in a region of elevated ionization parameter relative to the demarcation. While this result is suggestive of an AGN photoionization interpretation, we caution the limitation of the OHNO diagram to identify the source of ionization. In particular, as noted by other studies (e.g., \citealt{Larson2023, Calabro2024, Rinaldi2025}), star-forming photoionization models with low metallicity and high ionization parameter can be pushed toward the same region. However, low metallicity and high ionization parameter cannot account for the elevated electron temperatures probed by the O3H$\gamma$--Ne3O2 and O3H$\gamma$--O33 diagrams, as shown above, where LRDs lie beyond the range spanned by stellar photoionization models.

\subsubsection{O1 diagram}
The O1 diagram \citep{Veilleux1987} uses the line ratios [O\,\textsc{iii}]\,$\lambda$5007/H$\beta$ and [O\,\textsc{i}]\,$\lambda$6300/H$\alpha$. Similar to OHNO, this diagnostic relies on discriminating between high and low ionization conditions. 
At high-redshift, this diagnostic still robustly separates AGN and star-forming galaxies as compared to the typical BPT diagram (e.g., \citealt{Mazzolari2024, Juodzbalis2026}). We show the demarcation line from \cite{Mazzolari2024} in Figure \ref{fig:line_diagnostic}. Among the six LRDs with sufficiently high S/N spectra, at least three have line ratios within the AGN locus.  
Indeed, this disfavours stellar photospheres as the ionizing source for these LRDs, which would otherwise show weaker [O\,\textsc{i}]/H$\alpha$ at a given [O\,\textsc{iii}]/H$\beta$. 
While the majority of the sample lacks the sensitivity needed to exploit this diagnostic, the elevated [O,\textsc{i}] emission in those observed points to an extended partially ionized region in LRDs that requires either shock excitation or a hard ionizing continuum from an AGN.

\subsubsection{He\,\textsc{ii}\,$\lambda$4686/H$\beta$} 
The presence of high-ionization emission lines would also provide another clue for the presence of AGN. Lines such as C\,\textsc{iv}\,$\lambda$1549, N\,\textsc{v}\,$\lambda$1240, [Ne\,\textsc{v}]\,$\lambda$3426, and He\,\textsc{ii}\,$\lambda$4686 have ionization potentials that are $\gtrsim50$ eV, and are rarely prominent in star-forming regions, such that the resulting line fluxes relative to the Balmer lines are generally much weaker than what is observed in AGN. Given that the majority of the LRDs in our sample have line ratios consistent with AGN photoionization, we test whether those LRDs have any detections of high-ionization lines.

In particular, we focus on He\,\textsc{ii}$\lambda$4686, as this line is the most accessible given the wavelength coverage of our spectra explored. He\,\textsc{ii} has an ionization potential of 54.4 eV, and the He\,\textsc{ii}/H$\beta$ ratio combined with the [N\,\textsc{ii}]/H$\alpha$ ratio (N2) is a good AGN diagnostic that is known to hold at high redshift (e.g., \citealt{Ubler2023, Mazzolari2025}). However, as N2 is usually not robustly detected in LRDs, we instead adopt a fiducial threshold of $\log(\mathrm{He\,\textsc{ii}/H\beta}) = -1.22$ based on the asymptotic behaviour of the demarcation line defined by \cite{Shirazi2012}, above which a source can no longer be explained by standard star-formation driven photoionization. Of the three LRDs with He\,\textsc{ii} detections, one exceeded this threshold, albeit with a lower value compared to what is typically observed in local AGNs \citep{Shirazi2012, Tozzi2023}. If the emission lines of LRDs are powered by an AGN, these low observed He\,\textsc{ii}/H$\beta$ ratios suggest that LRDs have intrinsically weak high-ionization line emission above 54 eV, broadly consistent with previous findings for JWST-selected AGN (e.g., \citealt{Wang2025, Zucchi2026, Brazzini2026}).


\begin{table*}
\centering
\caption{Summary of narrow-line diagnostics for the LRD sample. A checkmark (\checkmark) indicates AGN-like line ratios and a cross (\ding{55}) indicates ratios consistent with star formation. A dash (--) indicates that the relevant lines were not detected at sufficient S/N (less than $3$), and the ellipsis (\nodata) indicates that the line lies outside the wavelength coverage.}
\label{tab:line_diagnostic}
\begin{tabular}{rrrrccccl}
\hline
PID & SRCID & RA & Dec & O3Hg/Ne3O2 & O3Hg/O33 & O3Hb/Ne3O2 & O3Hb/OHa & $\log$ He\,\textsc{ii}/H$\beta$ \\
\hline
1181 & 28074 & 189.064589 & 62.273815 & \checkmark & \ding{55} & \checkmark & \ding{55} & $-1.60^{+0.09}_{-0.12}$ \\
3215 & 162842 & 53.086501 & -27.891797 & \ding{55} & \ding{55} & \checkmark & -- & -- \\
3215 & 5070 & 53.092003 & -27.903136 & \nodata & \ding{55} & \nodata & -- & $-0.97^{+0.10}_{-0.16}$ \\
1286 & 38562 & 53.135864 & -27.871645 & \checkmark & \ding{55} & \checkmark & -- & -- \\
4106 & 47962 & 214.892479 & 52.856891 & -- & -- & \checkmark & -- & -- \\
1181 & 53501 & 189.295056 & 62.193572 & \checkmark & \ding{55} & \checkmark & \checkmark & -- \\
1181 & 73488 & 189.197396 & 62.177233 & \checkmark & \ding{55} & \checkmark & -- & -- \\
1181 & 68797 & 189.229137 & 62.146190 & \checkmark & \ding{55} & \checkmark & \ding{55} & -- \\
1180 & 13329 & 53.139038 & -27.784433 & \checkmark & \checkmark & \checkmark & -- & -- \\
4106 & 85168 & 214.830660 & 52.887775 & -- & \ding{55} & -- & \nodata & -- \\
3215 & 43564 & 53.044296 & -27.867604 & -- & -- & \checkmark & -- & -- \\
3215 & 642396 & 53.095910 & -27.906901 & -- & \ding{55} & -- & -- & -- \\
5224 & 920396 & 34.247828 & -5.151996 & -- & -- & \checkmark & -- & -- \\
4233 & 55604 & 214.983026 & 52.956001 & -- & \ding{55} & -- & \checkmark & -- \\
3215 & 160128 & 53.050301 & -27.900055 & -- & \ding{55} & -- & -- & -- \\
4106 & 57146 & 214.892246 & 52.877410 & \checkmark & \ding{55} & \checkmark & \checkmark & $-1.36^{+0.10}_{-0.12}$ \\
1286 & 159717 & 53.097528 & -27.901260 & -- & -- & \checkmark & \ding{55} & -- \\
4106 & 83161 & 214.809150 & 52.868480 & \nodata & \ding{55} & \nodata & \nodata & \nodata \\
\hline
\end{tabular}
\raggedright
\tablenotetext{a}{Sources are classified as AGN if $\log$(He\,\textsc{ii}/H$\beta$) $> -1.22$.}
\end{table*}

\begin{figure}[!t]
    \centering
    \includegraphics{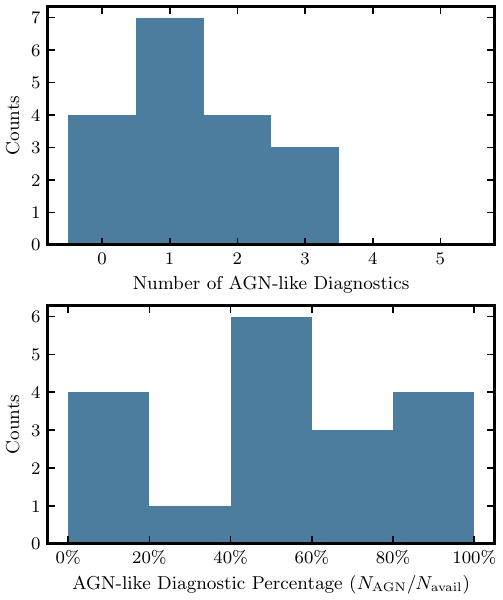}
    \caption{Top panel: Number of diagnostics identifying each LRD as an AGN. Bottom panel: The percentage of available diagnostics that classify each LRD as AGN-like. We find that at least half of LRDs show multiple line diagnostics consistent with AGN photoionization.
    }
    \label{fig:line_diagnostic_combined}
\end{figure}

\subsection{Are the Line Diagnostics Consistent?}
Given the range of results from different diagnostic tools, we now investigate whether they provide consistent classifications. The diagnostics are designed to probe the electron temperature and ionizing condition of the emitting gas, where objects found within the AGN parameter space are suggested to be subjected to more extreme ionizing conditions. While diagnostics such O3H$\beta$--Ne3O2 and O3H$\gamma$--O33 may be more ambiguous due to the overlap between SFGs and AGNs, finding consistency across multiple diagnostics would provide stronger support for the AGN interpretation. We summarize the line diagnostics results where emission lines are robustly detected in Table \ref{tab:line_diagnostic}.  

In Figure \ref{fig:line_diagnostic_combined}, the top panel shows the number of line diagnostics classifying each LRD as AGN-like. At least 40\% of the LRDs have more than 2 diagnostics that are AGN-like. The bottom panel shows the percentage of available diagnostics that classify each LRD as AGN-like. The distribution is skewed toward higher percentages ($>50\%$), suggesting that when multiple diagnostics are available, they broadly agree in favouring of an AGN-like ionizing source. We note here that some LRDs only have one diagnostic available, however, this only affects the tails of the distribution shown in the bottom panel. Restricting to LRDs with at least two diagnostics does not change this interpretation.

\begin{figure}[!t]
    \centering
    \includegraphics{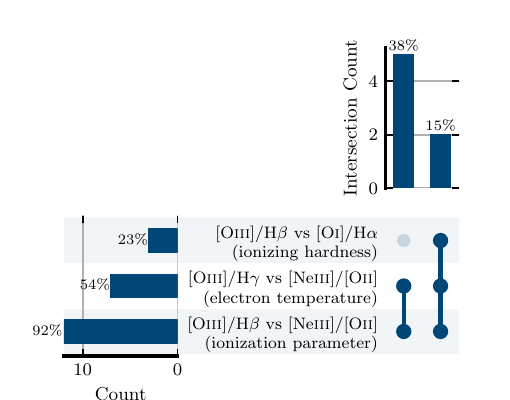}
    \caption{An UpSet diagram showing how the AGN classifications from different line diagnostics overlap. Here non-detections are treated as non-AGN, so the percentages represent lower limits. The bar plot on the top right shows the number of LRDs with overlapping diagnostics. At least 38\% of LRDs have line ratios consistent with high ionization parameter and electron temperature but with a non-AGN-like ratio for the O1 diagnostic. At least 15\% of LRDs in the sample have all three diagnostics collectively pointing to AGN or shocks. The bar plot on the bottom left shows the number of LRDs classified as AGN-like by each diagnostic individually.
    }
    \label{fig:upset_diagnostics}
    
\end{figure}

We also further explore how the AGN classifications from different line diagnostics overlap using an Upset diagram\footnote{\href{https://github.com/jnothman/UpSetPlot}{https://github.com/jnothman/UpSetPlot}} as shown in Figure \ref{fig:upset_diagnostics}. Here, we limit the analyses to the O3H$\gamma$--Ne3O2, O3H$\beta$--Ne3O2, and O3H$\beta$--O\textsc{i} diagnostics, omitting the He\,\textsc{ii}-based diagnostics due to mostly non-detections and the O33 diagnostic due to the difficulty of interpreting its line ratios as noted in \S \ref{sec:auroral_line}. For these three diagnostics, we treat non-detections as non-AGN, so the quoted percentages represent lower limits for LRDs with AGN-like line ratios. We find that at least 38\% of LRDs have line ratios consistent with high ionization parameter (based on O3H$\beta$--Ne3O2) and high electron temperature (based on O3H$\gamma$--Ne3O2) but with a non-AGN-like ratio for the O1 diagnostic. Furthermore, a non-negligible fraction of at least 15\% have all three diagnostics pointing to high ionization parameter, high electron temperature, and harder ionizing radiation, indicative of mechanisms such as shocks and AGN activities at plays. 

\section{Discussion \& Conclusion} \label{sec:discussion}

LRDs exhibit broad emission lines typically associated with AGN, yet other characteristic signatures such as X-ray emission and high-ionization lines are largely absent (e.g., \citealt{Lambrides2024, Tang2025}). Explaining these observations leads to a framework where the central engine is surrounded by dense neutral gas. This gas envelope would be responsible for both the Balmer break and absorption features (e.g., \citealt{Inayoshi2025}), as well as the suppression of X-ray emission \citep{Juodzbalis2024}.
It would reprocess high energy photons, resulting in an envelope with a characteristic effective temperature of $T_\mathrm{eff}=$ 5000 K (e.g., \citealt{Kido2025, Inayoshi2025, Begelman2026, Graaff2025_bh, Torralba2026}). 
Such a cool photosphere would produce a stellar-like ionizing spectrum, with predictable implications for the narrow-line diagnostics of LRDs, where line ratios should be inconsistent with AGN-like photoionization. In this work, we therefore utilize multiple different line diagnostics to place constraints on the ionization source of LRDs.

\subsection{What drives the narrow-line ratios?}
We find that at least 40\% of the LRDs in our sample have at least two narrow-line diagnostics that are consistently pointing toward ionization from an AGN (see Fig. \ref{fig:line_diagnostic_combined}), with extreme gas conditions atypical of star-forming galaxies. The line ratios indicate gas conditions associated high electron temperature and ionization parameter. Additionally, at least 15\% of LRDs also show elevated [O\,\textsc{i}]/H$\alpha$ ratios in the AGN regime, which are not typically produced by stellar populations. Geris et al. (in prep.) likewise report AGN-like line ratios in their sample of Little Red Dots and Little Blue Dots.

These classifications rest on several independent diagnostics. 
The [O\,\textsc{i}]-based VO diagram in Fig. \ref{fig:line_diagnostic} shows that at least half of the LRDs have elevated [O\,\textsc{i}]/H$\alpha$ ratios, consistent with an extended partially ionized zone characterized by hard photoionization. The temperature of the gas is probed through the [O\,\textsc{iii}] auroral line, and the O33 line ratio (with $\log([$O\,\textsc{iii}$]\,\lambda5007/[$O\,\textsc{iii}$]\,\lambda4363) < 1.5$), pointing towards high gas temperature. These line ratios are not reproduced by star-forming photoionization models across the range of metallicity and ionization parameter explored, which points tentatively toward a harder ionizing source. We note that the interpretation of the photoionization models in the O33 line diagnostic can be complicated if the gas density is multiphase as suggested by \citet{Deugenio2025}, who inferred the presence of higher electron densities at $n_e \gtrsim 10^6~\mathrm{cm}^{-3}$. This is significantly higher than the typically assumed $n_e \sim 10^3~\mathrm{cm}^{-3}$. At such densities, the [O\,\textsc{iii}]\,$\lambda5007$ transition becomes collisionally de-excited and suppresses $\mathrm{O33}$ independent of the ionization conditions (e.g., \citealt{Martinez2025}), complicating the use of O33 as a temperature diagnostic. Further constraints on the hardness of the ionizing spectrum can therefore come from the detection of high-ionization lines. 

Despite the consistency across multiple independent diagnostics pointing toward AGN photoionization, most LRDs lack strong high-ionization line emission, with He\,\textsc{ii}$\lambda$4686/$H\beta$ $\lesssim0.1$. In stacks of PRISM spectra, most LRDs lie within the AGN-dominated regions of the C\textsc{iii}] EW vs. C\textsc{iii}]/He\,\textsc{ii} parameter space, albeit, are near the boundary distinguishing AGN and star formation \citep{Perez2026}. Taken together, this is suggestive that LRDs must have a harder spectrum compared to star-forming galaxies at energies $>13.6$ eV (within the context of single-zone photoionization models at typical ISM densities), yet have a softer spectrum than standard AGN at energies $>47$ eV as evidenced by the general lack of C\textsc{iv} while C\textsc{iii}] is detectable.
Such a softer ionization spectrum could be attributed to super-Eddington accretion (e.g., \citealt{Pacucci2024, Lambrides2024}). We further note that Wolf-Rayet stars can also produce He\,\textsc{ii}, seen as a ``blue bump" feature around $\lambda4650~\text{\AA}$ \citep{Brinchmann2008}, and therefore He\,\textsc{ii} alone may not be a definitive tracer of the ionizing spectrum without supporting constraints from other rest-frame UV lines. Alternatively, the gas envelope invoked to explain the Balmer break and X-ray weakness or the surrounding accretion disk could also preferentially attenuate higher energy photons (e.g., \citealt{Madau2025}), producing an effectively softer spectrum in the narrow-line region.  

\begin{figure*}[t!]
    \centering
    \includegraphics[width=\textwidth]{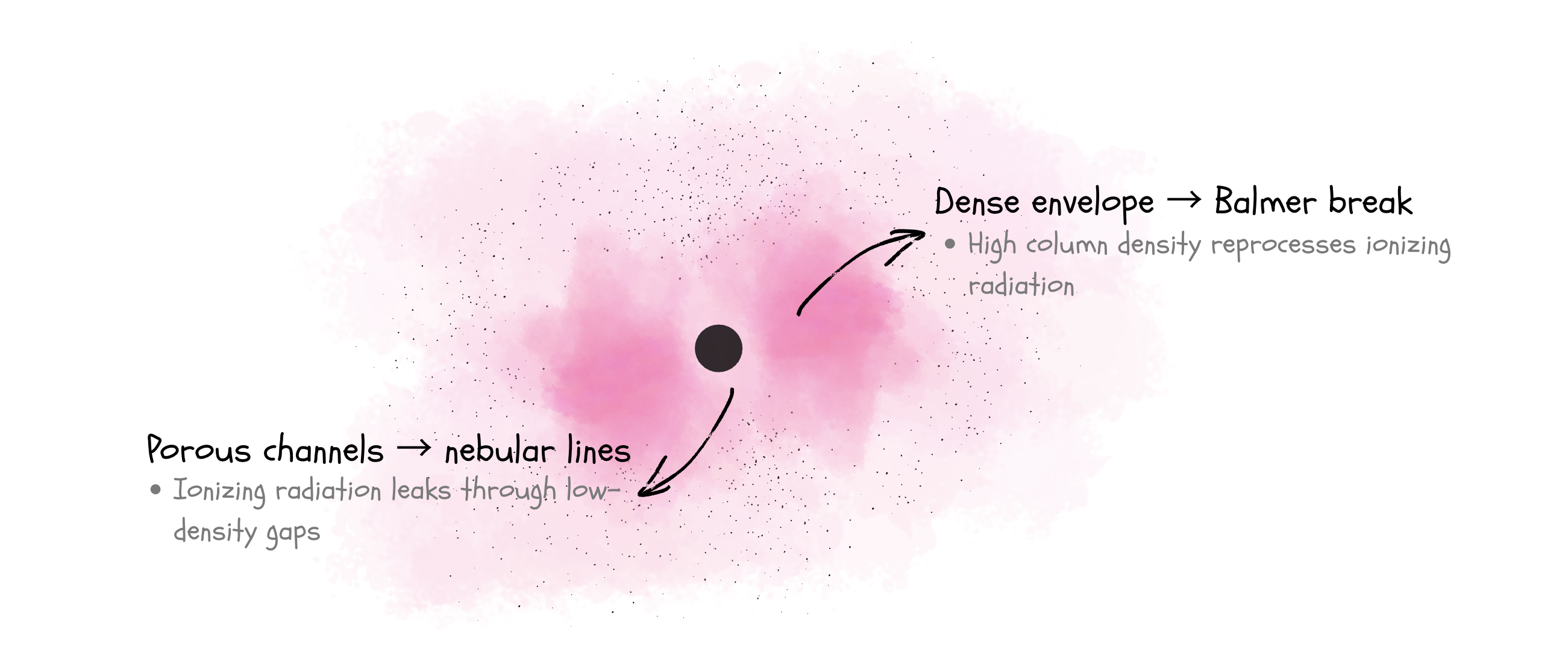}
    \caption{A cross section of the possible gaseous structure surrounding LRDs. Incident radiation that goes through the higher gas columns will result in the Balmer break and absorption features observed in LRDs. The AGN-like line ratios suggest that there are low-density channels, possibly analogous to funnel-like regions, where the incident radiation is not reprocessed.}
    \label{fig:schmatic}
\end{figure*}

We also consider whether shocks contribute to the observed line ratios as shocks can raise electron temperatures without producing extremely high ionization states. They could arise from AGN or stellar feedback, but may also trace the relatively slow winds of a few hundred km s$^{-1}$ predicted in quasi-star models \citep[e.g.,][]{Hassan2026, Santarelli2026}. 
Fully disentangling contributions from star formation, AGN, and shocks in galaxies generally require IFU data to combine spatially resolved line diagnostics with velocity dispersion maps \citep[e.g.,][]{Zhu2025}. 
Nevertheless, the intrinsic narrow-line full widths at half maximum in our sample have a median of \nfwhmme, with a 16th--84th percentile range of \nfwhmlo to \nfwhmhi, and are consistent with typical values reported in high-redshift AGNs \citep[e.g.,][]{Alvarez2025, Mazzolari2025}. 
This implies that the observed line ratios and absence of high-ionization emission are unlikely to be driven primarily by shocks.

\subsection{Implications on the gas geometry}
Our findings suggest that there must be lower density channels within the gas envelope in at least some LRDs where ionizing radiation from an AGN can escape. A similar conclusion was suggested in \cite{Matthee2026}, who attributed the correlation between the Balmer break and absorber velocities to outflowing material along these low-density channels. Additional evidence comes from recent observational constraints on the Ly$\alpha$ line profile in LRDs. In general, Ly$\alpha$ is detected in at least 40\% of LRDs \citep{Asada2026}, typically with low EW \citep{Ando2026uvlittlereddots}. The absence of broad Ly$\alpha$ emission in LRDs places constraints on the gas columns, pointing to a covering fraction close to unity \citep{Torralba2026_lyman}. In contrast, detections of broad Ly$\alpha$ emission cannot be explained by high gas columns with near-unity covering fraction, and instead suggest a gas medium that is clumpy, or porous \citep{Tang2026, Ji2026}. Together, these observations point to the presence of low-density channels in at least some LRDs.

We show a possible representation of the gas geometry surrounding LRDs in Figure~\ref{fig:schmatic}. Within the dense gas envelope, there exist low-density channels where ionizing photons can escape, potentially analogous to the funnel-like regions in super-Eddington accretion models (\citealt{Pacucci2024, Madau2025}). Photons passing through higher gas column densities produce a reprocessed continuum with a prominent Balmer break and redder optical slope. Balmer absorption features may arise when viewing LRDs along sightlines that pass through outflowing material (e.g., \citealt{Matthee2026}). 
The occasional detection of high-ionization lines in some LRDs (e.g., \citealt{Tang2025}), as well as the recent discovery of an X-ray emitting LRD \citep{Hviding2026}, may suggest a viewing angle directly into the funnel in super-Eddington accretion models, where radiation escaping along the funnel axis is significantly harder and more luminous, while extreme UV and soft X-ray output is strongly suppressed along equatorial sightlines \citep{Madau2025, Madau2026_unified}. We note, however, that at least one dynamical mass measurement of an LRD places the black hole below the Eddington limit \citep{Juodzbalis2025}, suggesting that super-Eddington accretion may not be universal among LRDs.

While we cannot determine the specific configuration of the dense gas envelope from these data alone, it is clear that the analyses of the line diagnostics show that a non-negligible fraction of at least 15\% LRDs have line ratios that are incompatible with star-forming photoionization models. This disfavours a uniform dense gas envelope, which would produce stellar-like line ratios, and is indicative of low-density channels existing within the gas envelope through which ionizing radiation from the AGN can escape. Future deep high spectral-resolution observations of the rest-frame UV will be valuable to constrain both key density-tracer lines and high-ionization lines, which will offer insights into the interstellar medium of the narrow-line emitting regions and further place crucial constraints the gas geometry of LRDs.

\begin{acknowledgments}
This work is based on observations made with the NASA/ESA/CSA James Webb Space Telescope. The data were obtained from the Mikulski Archive for Space Telescopes at the Space Telescope Science Institute, which is operated by the Association of Universities for Research in Astronomy, Inc., under NASA contract NAS 5-03127 for JWST. These observations are associated with program ID 4106 and 3215, along with 1181, 1286, 4233, 5224 from the archive. 
 
EJN acknowledges support for JWST-GO-03383, provided by NASA through a grant from the Space Telescope Science Institute. 
FDE acknowledges support by the Science and Technology Facilities Council (STFC), by the ERC through Advanced Grant 695671 ``QUENCH'', and by the UKRI Frontier Research grant RISEandFALL.
AJB acknowledges funding from the “FirstGalaxies” Advanced Grant from the European Research Council (ERC) under the European Union’s Horizon 2020 research and innovation program (Grant agreement No. 789056). H\"U acknowledges support by the Max Planck Society through the Lise Meitner Excellence Program. H\"U acknowledges funding by the European Union (ERC APEX, 101164796). Views and opinions expressed are however those of the authors only and do not necessarily reflect those of the European Union or the European Research Council Executive Agency. Neither the European Union nor the granting authority can be held responsible for them.

Some of the data products presented herein were retrieved from the Dawn JWST Archive (DJA). DJA is an initiative of the Cosmic Dawn Center (DAWN), which is funded by the Danish National Research Foundation under grant DNRF140.

\end{acknowledgments}

\bibliography{article}{}
\bibliographystyle{aasjournalv7}



\end{document}